\documentclass{emulateapj}
\usepackage{wasysym}
\usepackage{natbib}
\usepackage{etoolbox}

\newtoggle{emulateapj}
\toggletrue{emulateapj}
\iftoggle{emulateapj}{\newcommand{\Sigurdsson}{Sigur{\scriptsize \DH}sson}}{\newcommand{\Sigurdsson}{Sigur\dh sson}}
\iftoggle{emulateapj}{\newcommand{\icarus}{Icarus}}{}

\begin{document}

\title{Earthshine on a Young Moon: \\Explaining the Lunar Farside Highlands}
\author{Arpita Roy, Jason T. Wright, \& Steinn \Sigurdsson}
\affil{Department of Astronomy \& Astrophysics, 525 Davey Lab, The Pennsylvania State University, University Park, PA, 16802}
\affil{Center for Exoplanets and Habitable Worlds, 525 Davey Lab, The Pennsylvania State University, University Park, PA, 16802}

\begin{abstract}
The lunar farside highlands problem refers to the curious and unexplained fact that the farside lunar crust is thicker, on average, than the nearside crust. Here we recognize the crucial influence of Earthshine, and propose that it naturally explains this hemispheric dichotomy. Since the accreting Moon rapidly achieved synchronous rotation, a surface and atmospheric thermal gradient was imposed by the proximity of the hot, post-Giant-Impact Earth.  This gradient guided condensation of atmospheric and accreting material, preferentially depositing crust-forming refractories on the cooler farside, resulting in a primordial bulk chemical inhomogeneity that seeded the crustal asymmetry. Our model provides a causal solution to the lunar highlands problem: the thermal gradient created by Earthshine produced the chemical gradient responsible for the crust thickness dichotomy that defines the lunar highlands.
\end{abstract}

\section{Introduction}

The lunar farside highlands problem refers to the fact that the farside lunar crust is thicker, on average, than the nearside crust, and presents a challenge to the current understanding of lunar formation and evolution. Within the resolution to this problem lies concrete knowledge of the Moon's assembly, an understanding of the solidification histories of planetary bodies, and insight relating to the geology of hot exoplanets that are close to their host stars. The Moon exhibits a dramatic dichotomy between hemispheres, especially in terms of topography \citep{Kaula:1974}, compositional variation \citep{Wieczorek:2013,  Jolliff:2000}, the ubiquity of volcanic maria \citep{Head:1992}, and crustal thickness \citep{Ishihara:2009,Wieczorek:2013,Zuber:1994}.  

Several mechanisms have been proposed as the origin of the lunar disparity, including those invoking external events such as the accretion of a companion moon \citep{Jutzi:2011}, asymmetric nearside-farside cratering \citep{Wood:1973}, and consequences of large impacts that formed the Procellarum \citep{Nakamura:2012} and South Pole-Aitken basins \citep{Zuber:1994}.  Others have explored internal phenomena, like asymmetric crystallization of the magma ocean \citep{Ohtake:2012,Wasson:1980}, tilted convection \citep{Loper:2002}, and spatial variations in tidal heating \citep{Garrick-Bethell:2010}. In reminiscence of rocky planets whose proximity to their stars dramatically affects their geologies \citep{Leger:2011}, here we recognize the crucial influence of Earthshine, and propose that the hemispheric dichotomy in crustal thickness emerges as a direct consequence of the conditions presiding over moon formation. 

The leading theory of lunar origin, broadly consistent with both dynamical and chemical constraints, is via giant impact between a Mars-sized impactor and the proto-Earth during its final stages of accretion \citep{Hartmann:1975, Cameron:1976}.  Regardless of the details of the progenitor masses, impact parameters, and relative velocity, this highly energetic collision is predicted to have melted and partially vaporized the impactor and large regions of the terrestrial mantle, sequestering the core material of both bodies into Earth, while iron-poor silicate material spun into a circumterrestrial disk that coalesced to form the Moon \citep{Canup:2004a,Canup:2001,Ida:1997}. Immediately after the collision, temperatures on Earth would have risen to $T_\oplus \sim$ 8,000 K \citep{Canup:2004}, which could only radiatively cool to $\sim$2,500 K, as Earth's atmosphere became defined by incandescent silicate clouds for the next thousand years \citep{Pahlevan:2007, Zahnle:2007}. Thus, to the forming Moon, the post-impact Earth would have been a proximate and strongly radiating presence that could greatly influence the course of its accretion.

\section{The Moon Formed Tidally Locked}

Our hypothesis requires that the Moon formed effectively tidally locked. The Moon is believed to have accreted rapidly just beyond the Roche limit \citep{Kokubo:2000}. The forming Moon would have experienced strong tidal damping and quickly evolved into a state of synchronous rotation with Earth, making it acceptable to assume that the Moon has always been tidally locked \citep{Stacey:1992, Peale:1978}. This can be verified using the tidal spin-down time \citep{Peale:1977}, 
\begin{equation}
\tau_{\rm spin} \sim Q \left(\frac{R_{\leftmoon}^3}{GM_{\leftmoon}}\right) \omega_{\leftmoon} \left(\frac{M_{\leftmoon}}{M_\oplus}\right)^2\left(\frac{D}{R_{\leftmoon}}\right)^6
\end{equation}
where $Q$ is the tidal dissipation factor, $\omega_{\leftmoon}$ is the Moon's primordial rotation rate, and $D$ is the Roche radius $(\sim 3 R_\odot)$.  Using the average contemporary $Q$ of 35 \citep{Williams:2012}, and a lower limit to the initial spin period of 1.8 hours \citep{Garrick-Bethell:2006} shows that the Moon would have definitely tidally locked by $\sim$~100~days. Since the low estimate of the spin period is set in the limit of rotational instability, it is likely that the Moon locked even earlier, and certainly before the last stages of accretion, which takes 1-$10^2$ years \citep{Salmon:2012}. This is also orders of magnitude quicker than the crust formation timescale, which is defined by the idea that the lunar magma ocean solidified to 80\% in $10^3$ -- $10^4$ years, after which the flotation of plagioclase slowed down the cooling (and hence solidification) rate further \citep{Meyer:2010}. The subsequent development of a thicker farside crust would strengthen the locked configuration as the Moon cooled \citep{Loper:2002}.  Any episodes of asynchronicity that occurred after crust-formation (caused, for instance, by later impacts) have no bearing on our model, as long as the Moon ultimately regained its original (and current) configuration, consistent with simulations that suggest that the current configuration has always been favored \citep{Aharonson:2012}.

\section{Effects of Earthshine on the Moon}

The effects of Earthshine on the tidally-locked infant Moon have never been adequately acknowledged in previous studies of lunar formation and evolution. The only prior calculation of this temperature gradient assumed that the only effect of temperature asymmetry was to delay crystallization, and suggested a difference of only 11K between hemispheres \citep{Wasson:1980}. But this calculation assumed equal cooling times for Moon and Earth, and was valid only after both had cooled to $\sim$850 K (i.e., long after accretion was complete and the crust had begun to form). However, the Moon has a larger surface area to volume ratio and cooled much faster than Earth.  At earlier times, the Moon's nearside would have been continuously irradiated by Earth while the farside remained in the dark, setting up a major surface temperature asymmetry. Although both Earth and Moon start out very hot, the lunar farside equilibrium temperature is set by the Sun,

\begin{equation}
  T_{\rm far} = \left[\frac{T^4_\odot R^2_\odot}{4 D_\odot^2}\right]^{1/4}
\end{equation}

while the nearside equilibrium temperature is dominated by the much brighter contribution from Earthshine,

\begin{equation}
  T_{\rm near} = \left[\frac{T^4_\odot R^2_\odot}{4 D_\odot^2} + \frac{T^4_\oplus R^2_\oplus}{2D^2_\oplus}\right]^{1/4}
\end{equation}

\noindent where $T$ is temperature, $R$ is radius, and $D$ is distance from the moon.  When $D_\oplus = 3 R_\oplus$, Earth would have angular diameter of $40^\circ$, occupying 7\% of the lunar sky.  Thus, the farside attempts to cool towards an effective temperature of $\sim 250$~K set only by the solar flux, while the closest point to Earth cannot cool below $(0.07)^{1/4} T_\oplus\sim T_\oplus /2$ due to thermal insulation by Earth.  The dominance of Earthshine is not surprising, given that the geometry and temperatures of the proto-Earth-Moon system are comparable to those of close-in exoplanets orbiting $\sim 3$ stellar radii from K and M dwarf stars. The large difference in these equilibrium temperatures implies a much higher cooling rate for the farside; thus, a strong nearside-farside temperature gradient is established as soon as the proto-Moon is largely assembled and begins to cool, even if neither hemisphere actually achieves its equilibrium temperature during the last stages of accretion.

Our hypothesis is only valid if the cooling time of the lunar atmosphere and disk is shorter than the Moon's accretion timescale and Earth's cooling timescale. The accretion of the Moon takes 1-$10^2$ years, with a proposed early rapid phase ($\sim0.1$ years) in which material initially outside the Roche limit coalesces through impacts, and a protracted phase ($\sim10^2$ years) in which material is delivered to the outer disk as the Roche-interior disk viscously spreads \citep{Ida:1997,Kokubo:2000,Salmon:2012}. Earth's surface remains at 2,000 -- 10,000 K for $\sim$ 2,000 yr, similar to the timescale for plagioclase flotation.  Even while the Moon is being assembled, however, the material in the lunar atmosphere and protolunar disk that will form the Moon has time to cool significantly as the second phase of accretion progresses. The cooling time for the lunar atmosphere and protolunar disk can be estimated to a rough order of magnitude as
\begin{equation}
  \tau_{cool} \sim \frac{C_p m T}{4 \pi \sigma_{\rm SB} T_{\leftmoon}^4 R_{\leftmoon}^2}
\end{equation}
where $m$ is the mass of the radiating lunar atmosphere and surface (on the order of 1\% of the Moon's total mass),  $\sigma_{\rm SB}$ is the Stefan-Boltzmann constant, and the specific heat is $C_p \sim 10^7$ erg g$^{-1}$K$^{-1}$. Thus, the lunar atmosphere and protolunar disk cool rapidly, on the order of 1 yr, and in the absence of heating would approach their equilibrium temperatures, set primarily by Earthshine while accretion is still ongoing. The temperature gradient is already established as the later stages of lunar assimilation proceed, and it consequently influences the deposition pattern.

\section{Effects of Thermal Gradient on Young Moon Composition}

The earliest lunar epoch, when the last $\sim$10\% of the lunar mass was still being assembled, was characterized by widespread magma oceans that acted as sources of rock vapor for a thick primordial lunar atmosphere \citep{Stern:1999} and an extended protolunar disk. For example, a bulk silicate oxide (SiO$_4$) atmosphere would have had a scale height $H \sim$ 200 km.  In this massive, dynamic atmosphere, characterized by high temperatures and large amounts of stochastic, localized heating from accertion events, significant amounts of both refractories and volatiles would have been in gas, liquid, and solid phases \citep{Elkins-Tanton:2013}. 
 
As the  nearly-completed Moon cooled, condensation would necessarily have been guided by the local temperature gradient.  Regardless of the specific temperature and pressure at which the various refractory species rained out of the protolunar atmosphere, the farside of the Moon, having cooled first, would lead the nearside in condensation.  Condensation in the part of the protolunar disk shaded from Earthshine by the Moon's shadow may have also played a role in preferentially delivering refractories to the farside.  The bulk abundance of refractory species (led by Al$_2$O$_3$ and CaO, the major gases of Al and Ca) was thus increased in the farside melt.

We couple this to the observation that the present-day crust is indeed composed of Ca- and Al-enriched silicate minerals \citep{Taylor:2009}. The highlands crust is anorthositic with a high content of plagioclase ($\sim 90\%$) and high Al$_2$O$_3$ concentrations \citep{Taylor:2009}. In fact, the Al$_2$O$_3$ content of the magma ocean (estimated as ~4 wt.\% on average) is often used to determine the thickness of crust it would produce \citep{Elkins-Tanton:2011}. Thus, an early enrichment of crust-forming aluminum and calcium-bearing refractories in the farside magma ocean would have naturally enhanced plagioclase formation in that region.  In this scenario it is not necessary to invoke any extreme internal redistribution of lunar material to provide the excess of Ca and Al condensates required to form a thicker crust on the farside. Any hemispheric variations in Ca and Al enriched material would be reflected in the regional crustal thickness, and conversely, any crustal thickness dichotomy speaks to a chemical dichotomy in plagioclase forming condensates. Essentially, we argue that the chemical dichotomy implied by the variation in lunar crustal thickness is primordial, and the Moon's primordial temperature dichotomy naturally explains the consequent compositional gradient.

Our hypothesis is only valid if the majority of crust formation occurred in the presence of a hemispheric chemical gradient, before significant azimuthal mixing due to convection. The patterns of early lunar convection are consequently an important consideration, since adequate mixing prior to crust formation could eradicate the chemical gradient and invalidate the possibility that the farside highlands are the result of a primordial signature. Note that in our model it is not essential for the veneer of the primary crust to be retained on the surface (for example, after cumulate mantle overturn), only that the inhomogeneity in low-order azimuthal refractory distribution established via the temperature gradient is maintained.  Subsequent impacts and local mixing do not affect our model, since they will not transport the bulk of the crust-forming deposits to the nearside.
 
Unfortunately, the nature of the convection in these early stages of lunar accretion is not well established. A simple calculation using lunar magma ocean parameters from \citet{Suckale:2012}, which do not take into account the differences between planetary-scale thermal convection and composition-driven local convection, suggests that the characteristic turnover time of the lunar magma ocean is 3 hr - 30 yr.  Considerable uncertainty remains, however, in the form and scale of this convection, and therefore in how efficiently it would mix the magma ocean on global scales.  Numerical simulations of a more evolved magma ocean, that include crystallization products and account for both compositional and thermal buoyancy effects, indicate a characteristic mixing time of $\sim$200 Myr \citep{Spera:1992}, and \citep{Elkins-Tanton:2011} argued that even over millions of years, compositions are not likely to have been thoroughly mixed.  Obviously, the fact that the Moon is observed to have compositional variations \citep{Wieczorek:2013, Jolliff:2000} suggests that convection in the magma ocean or mantle has not been efficient enough to create complete global chemical homogeneity.

\section{Conclusion}

The thermodynamic and physical conditions that presided over the formation of the Moon under the Giant Impact Hypothesis are highly uncertain \citep{Elkins-Tanton:2013}. We find that the dichotomy in lunar crustal thickness \citep{Ishihara:2009} helps to constrain these conditions if the chemical inhomogeneity observed in the lunar hemispheres is a relic of an early temperature gradient on the tidally-locked Moon, acting in concert with thermal variations in the protolunar disk and atmosphere, due to Earthshine.  An advantage of our model over theories that invoke largely stochastic accretion or displacement of lunar material  is that it provides a more deterministic explanation for the farside highlands.  While more detailed studies are necessary to substantiate our model, our work here underlines the need to include Earthshine in Moon-formation models and simulations, and provides a framework for future work in today's climate of investment into lunar formation.   If the deposition of refractory materials according to the thermal pattern was indeed an important aspect of the Moon's construction, then there may be observational consequences beyond the crust thickness dichotomy that studies comparing the hemispheres, such as that of \citet{Ohtake:2012}, may reveal.

\acknowledgements
We thank the following for helpful conversations: Neyda Abreu, Lynn Carter, Matija \'Cuk, Bethany Ehlmann, Andrew Ingersoll, James Kasting, David Stevenson, Yuk Yung, Gary Glatzmaier, and Kevin Zahnle. We thank our referees for their useful feedback.  The Center for Exoplanets and Habitable Worlds is supported by the Pennsylvania State University, the Eberly College of Science, and the Pennsylvania Space Grant Consortium. We also acknowledge support from the NASA Astrobiology Institute and the Pennsylvania State Astrobiology Research Center under grant number NNA09DA76A.


\end{document}